\documentclass[aps, twocolumn, prl, longbibliography]{revtex4-1}
\usepackage{mathptmx} % use Times font instead of Computer Modern
\usepackage{amsfonts}
\usepackage{amsmath}
\usepackage{amssymb}
\usepackage{graphicx}
\usepackage{bm}
\usepackage{natbib}
\usepackage{color}
\usepackage[colorlinks=true,linkcolor=blue,anchorcolor=red,citecolor=blue, urlcolor=blue]{hyperref}

\begin{document}
\title{Counterflowing edge current and its equilibration in quantum Hall  devices with sharp edge potential: Roles of incompressible strips and contact configuration}
\author{T. Akiho, H. Irie, K. Onomitsu, and K. Muraki}
\affiliation{NTT Basic Research Laboratories, NTT Corporation, 3-1 Morinosato-Wakamiya, Atsugi 243-0198, Japan}
\pacs{PACS number}
\date{\today}

\begin{abstract}
We report the observation of counterflowing edge current in InAs quantum wells which leads to the breakdown of quantum Hall (QH) effects at high magnetic fields.
Counterflowing edge channels arise from the Fermi-level pinning of InAs and the resultant sharp edge potential with downward bending.
By measuring the counterflow conductance for varying edge lengths, we determine the effective number $\langle N_\text{C} \rangle$ of counterflowing modes and their equilibration length $\lambda_\text{eq}$ at bulk integer filling factor $\nu = 1$--$4$.
$\lambda_\text{eq}$ increased exponentially with magnetic field $B$, reaching $200~\mu$m for $\nu = 4$ at $B \geq 7.6$~T.
Our data reveal important roles of the innermost incompressible strip with even filling in determining $\langle N_\text{C} \rangle$ and $\lambda_\text{eq}$ and the impact of the contact configuration on the QH effect breakdown.
Our results show that counterflowing edge channels manifest as transport anomalies only at high fields and in short edges.
This in turn suggests that, even in the integer QH regime, the actual microscopic structure of edge states can differ from that anticipated from macroscopic transport measurements, which is relevant to various systems including atomic-layer materials.
\end{abstract}
\maketitle

Understanding and controlling the electronic states at the edge of a two-dimensional system are becoming increasingly important.
This is particularly true for topologically nontrivial systems, such as quantum Hall (QH)~\cite{Halperin1982,Streda1987,Buttiker1988} and quantum spin Hall~\cite{Kane2005a,Kane2005} systems, where gapless edge states with distinct properties appear.
Recent theories~\cite{Lindner2012, Vaezi2013, Clarke2013, Mong2014, Clarke2014} predict that, by coupling their edge states to superconductors, QH as well as quantum spin Hall systems can be exploited to engineer exotic quasiparticles with non-Abelian statistics, a building block for robust quantum computation~\cite{Alicea2012, Leijnse2012}.
Semiconductor heterostructures comprising InAs, which can form transparent junctions with superconductors~\cite{Knez2012, Pribiag2015, DeVries2018}, are promising for such purposes.
Theory further predicts that certain fractional QH edge states coupled through a superconductor may harbor even more exotic quasiparticles that would allow for universal topological quantum computation~\cite{Lindner2012, Vaezi2013, Clarke2013, Mong2014, Clarke2014}.
Motivated by these predictions, recently the quality of InAs-based heterostructures has been improved significantly~\cite{Tschirky2017, Thomas2018}, which has led to the observation of a fractional QH effect~\cite{Ma2017}.

\begin{figure}[ptb]
\includegraphics{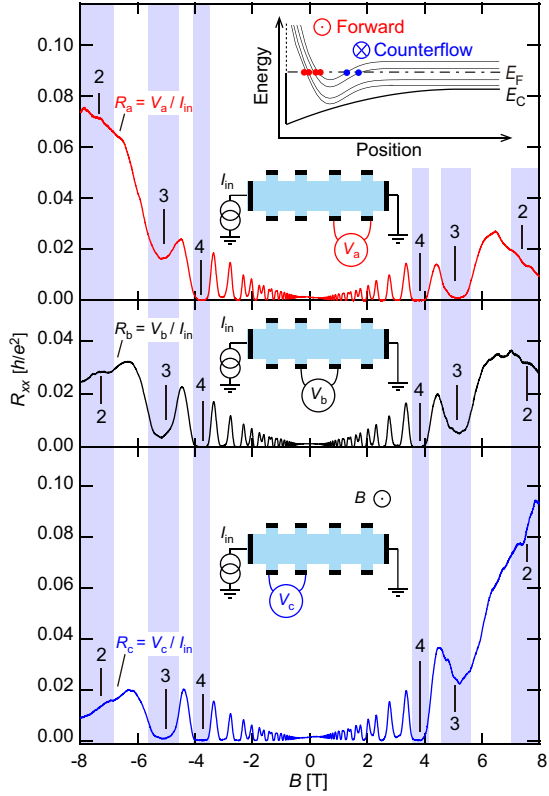}\caption{
Probe-position dependence of $R_{xx}$ vs $B$ of sample A.
The insets show the contact configuration for each measurement.
(Inset of upper panel) Schematic diagram of the conduction band edge ($E_\text{C}$) and Landau level dispersion in the presence of Fermi level pinning in the conduction band at the edge.
Edge channels are formed when the Fermi level ($E_\text{F}$) crosses Landau levels.}
\label{Fig1}
\end{figure}

In standard GaAs-based heterostructures, the edge potential is bent upward by the Fermi-level pinning in the band gap so that the electron density decreases monotonically toward the edge~\cite{Halperin1982, Chklovskii1992}.
This forms the basis for the common situation in QH systems where all edge channels have the same chirality, flowing in the same direction set by the magnetic field~\footnote{
In the fractional QH regime, electron correlation can lead to an edge state with a non-monotonic density profile and counterflowing modes, which is outside the scope of this work.
See Ref.~\cite{Lafont2019} and references therein.
}.
In contrast, in InAs the surface pinning occurs in the conduction band~\cite{Waldrop1984, Noguchi1991}, which implies that in heterostructures the edge potential is bent downward so the electron density increases near the edge.
While this is advantageous for superconducting junctions, it gives rise to trivial edge conduction with no topological origin at zero magnetic field~\cite{Nichele2016, Nguyen2016, Mueller2017, Mittag2017, DeVries2018}.
In a quantizing magnetic field, this suggests that the Fermi level can cross Landau levels extra times (see the inset to Fig.~\ref{Fig1}), where  additional sets of edge channels running in the forward and counterflow directions form~\cite{VanWees1995}.
As recently revealed in graphene~\cite{Cui2016}, a similar situation can also occur in a gated device due to electric-field focusing near the edge~\cite{Silvestrov2008}.

Counterflowing edge channels were first conceived by van Wees \textit{et al.}~\cite{VanWees1995}, who observed in their InAs quantum well that QH effects collapsed when a negative gate voltage below a certain threshold ($\sim -0.4$~V) was applied.
The results were then explained using the Landauer-B\"{u}ttiker model~\cite{Buttiker1988}, taking into account the scattering between forward and counterflowing edge channels, which indicated a typical equilibration length in excess of $200~\mu$m.
However, it remains unknown what determines the equilibration length and how it depends on the parameters such as the magnetic field and filling factor.
In this paper, we address these issues by systematically studying QH edge transport in InAs quantum wells using gated Hall-bar devices with only well-defined edges.
We directly detect the upstream charge current using a three-terminal setup, which allows us to determine the effective number of counterflowing modes and their equilibration length.
Our data reveal important roles of the innermost incompressible strip with even filling and the impact of the contact configuration for the counterflowing edge channels to manifest in transport.
Our results provide new insights into microscopic details of QH edge states, which will be useful for understanding edge transport in various systems including atomic-layer materials and in superconducting junctions, not only in the QH but also in the quantum spin Hall setups.

The heterostructure studied was grown by molecular beam epitaxy on an $n$-type GaSb (001) substrate.
The layer structure comprises a 20-nm-thick InAs quantum well sandwiched between Al$_{0.7}$Ga$_{0.3}$Sb barriers, with no intentional doping to supply carriers.
The center of the well is located $65$~nm below the surface of the 5-nm-thick GaSb cap.
The heterostructure was processed into $50$-$\mu$m-wide Hall bars as shown in the inset of Fig.~\ref{Fig1} by wet etching.
We fabricated devices with ten Ti/Au Ohmic electrodes and a Ti/Au gate on an atomic-layer-deposited 40-nm-thick Al$_{2}$O$_{3}$ insulator.
The gate covers all the mesa edges and their interface with Ohmic contacts, so that all the edges are defined in the same way.
The sample had sheet electron density of $n = 3.65 \times 10^{15}$~m$^{-2}$ and low-temperature mobility of $50$~m$^{2}$/Vs.
We used two samples fabricated from the same wafer, sample A with all edges having the same length of $L_\text{edge} = 60~\mu$m and sample B with varying $L_\text{edge}$ ($= 30$--$280~\mu$m).
Measurements were done at $1.5$~K using a standard lock-in technique.

We first present results for sample A.
Figure~\ref{Fig1} shows the magnetic field ($B$) dependence of the longitudinal resistance ($R_{xx}$) at front gate voltage $V_\text{FG} = 0$~V, measured using different pairs of voltage probes on the lower edge of the sample.
At $|B| \leq 4$~T, we observe normal behavior---Shubnikov-de Haas oscillations and well-developed QH effect at Landau-level filling factor  $\nu = 4$---for all configurations.
 ($\nu = nh/eB$ with $e$ the elementary charge and $h$ Planck's constant).
In contrast, anomalous behavior is seen at $|B| > 4$~T, where the QH effects expected at $\nu = 3$ and $2$ are not fully developed or completely missing, as seen by the non-vanishing $R_{xx}$.
Interestingly, the values of the finite $R_{xx}$ at $\nu = 3$ and $2$ systematically depend on the field direction and probe position.
At $\nu = 2$, $R_{xx}$ measured with the lower-right probes ($R_\text{a}$) is much higher for $B < 0$ than for $B > 0$.
Opposite behavior is seen for $R_{xx}$ measured with the lower-left probes ($R_\text{c}$), which is much higher for $B > 0$.
The lower-middle probes ($R_\text{b}$) gives intermediate values nearly symmetric for both field directions.
Although not shown, measurements using the probes on the upper edge confirm similar behavior, but with the probe-position dependence $180^\circ$ rotated around the sample normal.
We show below that this chiral breakdown behavior of the QH effect can be explained by the Landauer-B\"{u}ttiker model that takes into account the scattering between forward and counterflowing edge channels.

\begin{figure*}[ptb]
\includegraphics{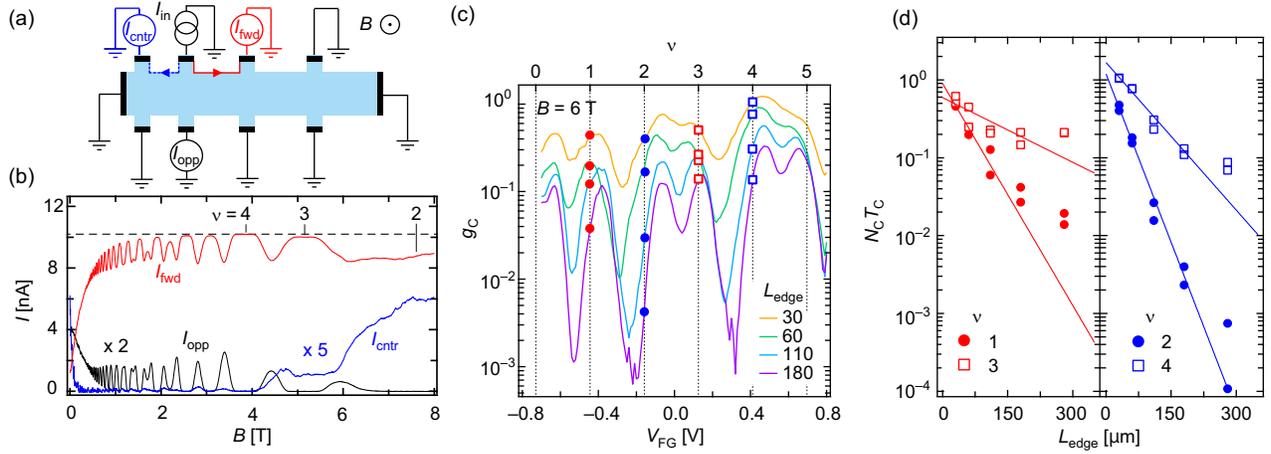}\caption{
(a) Schematic of the three-terminal measurement detecting upstream counterflowing current $I_\text{cntr}$ in addition to normal forward current $I_\text{fwd}$ (shown by the blue and red arrows, respectively).
Current $I_\text{opp}$ at the opposite side of the Hall bar was also monitored as a measure of bulk conduction.
(b) Magnetic-field dependence of $I_\text{fwd}$, $I_\text{cntr}$, and $I_\text{opp}$, measured in sample B at $V_\text{FG} = 0$~V using the configuration shown in (a).
(c) $V_\text{FG}$ dependence of the normalized counterflow conductance $g_\text{C}$ ($\propto I_\text{cntr}$) for different edge length ($L_\text{edge}$) at $B = 6$~T (see main text for details).
The top axis indicates the bulk filling factor estimated from the low-field Shubnikov-de Haas oscillations and Hall measurements at each $V_\text{FG}$.
(d) $L_\text{edge}$ dependence of $N_\text{C}T_\text{C}$ ($= g_\text{C}$) for $\nu = 1$--$4$ extracted from the data in (c).
Solid lines are fitting using a single exponential function.}
\label{Fig2}
\end{figure*}

We demonstrate the existence of counterflowing charge current using the three-terminal measurements as illustrated in Fig. 2(a), which in turn allowed us to directly determine the number of counterflowing modes ($N_\text{C}$) and their transmission probability ($T_\text{C}$) for individual edges.
In order to examine the $L_\text{edge}$ dependence of $T_\text{C}$, we used sample B with varying $L_\text{edge}$ ($= 30$--$280~\mu$m).
A magnetic field was applied in the direction so that the chirality of the edge channels was clockwise.
With this three-terminal setup, we detected charge current $I_\text{cntr}$ at the probe located on the upstream of the electrode from which current $I_\text{in}$ ($\sim 10$~nA) was driven, in addition to normal forward current $I_\text{fwd}$ measured on its downstream.
To check the conduction through the bulk, we also monitored current $I_\text{opp}$ on the opposite side of the Hall bar.
In the QH regime, where the current cannot flow through the bulk, the chirality requires $I_\text{fwd} = I_\text{in}$ and $I_\text{cntr} = 0$.
As shown in Fig.~\ref{Fig2}(b), we observe that this holds only at $B = 2$--$4$~T.
At $B > 4$~T, $I_\text{fwd}$ is seen to be noticeably lower than $I_\text{in}$ at fields where $I_\text{opp}$ is vanishing, accompanied by a significant increase in $I_\text{cntr}$.
This observation of upstream charge current in the QH regime provides direct evidence for the existence of counterflowing edge channels.

Using the voltage $V_\text{in}$ applied to drive $I_\text{in}$ and measured currents $I_\text{fwd}$ and $I_\text{cntr}$ in the QH regime, we define the conductance in the forward and counterflow directions as $g_\text{F}^{(i)} = (I_\text{fwd}/V_\text{in})/G_{0}$ and $g_\text{C}^{(i - 1)} = (I_\text{cntr}/V_\text{in})/G_{0}$ for the edges on the downstream and upstream labeled $i$ and $i-1$, respectively, in units of conductance quantum $G_{0} = e^{2}/h$.
In the Landauer-B\"{u}ttiker model~\cite{Buttiker1988, VanWees1995}$,  g_\text{C}$ can be expressed as $g_\text{C}^{(i-1)} = N_\text{C}T_\text{C}^{(i-1)}$, where $T_\text{C}^{(i-1)}$ is the transmission probability of the counterflowing mode of the edge on the upstream labeled $i-1$.
Note that there are $\nu + N_\text{C}$ forward edge channels in the presence of $N_\text{C}$ counterflowing edge channels.
Detailed balance requires  $g_\text{F}^{(i)} = \nu + g_\text{C}^{(i)}$ for each edge~\footnote{
Detailed balance requires $(\nu + N_\text{C})(1-T_\text{F}^{(i)}) = N_\text{C}(1-T_\text{C}^{(i)})$ for each edge, where $T_\text{F}^{(i)}$ is the transmission probability of the forward modes on the $i$th edge. By solving this for $T_\text{F}^{(i)}$ and pluging it into $g_\text{F}^{(i)} = (\nu+ N_\text{C})T_\text{F}^{(i)}$, we have $g_\text{F}^{(i)} = \nu+ N_\text{C}T_\text{C}^{(i)} = \nu+ g_\text{C}^{(i)}$.
}.
In what follows, we therefore show only results for $g_\text{C}$.
We repeated similar three-terminal measurements using the same sample while sequentially changing the injector and detector contacts, which allowed us to evaluate $g_\text{C}$ for different edges.
Figure~\ref{Fig2}(c) shows $g_\text{C}$ for different $L_\text{edge}$, obtained while sweeping $V_\text{FG}$ at a fixed magnetic field of $6$~T.
The top axis shows the bulk filling factor determined from the low-field Shubnikov-de Haas oscillations and Hall measurements at each $V_\text{FG}$.
We note that $g_\text{C}$ oscillates with $V_\text{FG}$, but with the positions of the minima shifted from the bulk integer filling to lower $V_\text{FG}$~\footnote{
The fact that $g_\text{C}$ decreases as $V_\text{FG}$ is slightly lowered from integer $\nu$ is consistent with the conjecture that the electron density in the vicinity of the mesa edge is higher than that in the bulk.}

\begin{figure}[b]
\includegraphics{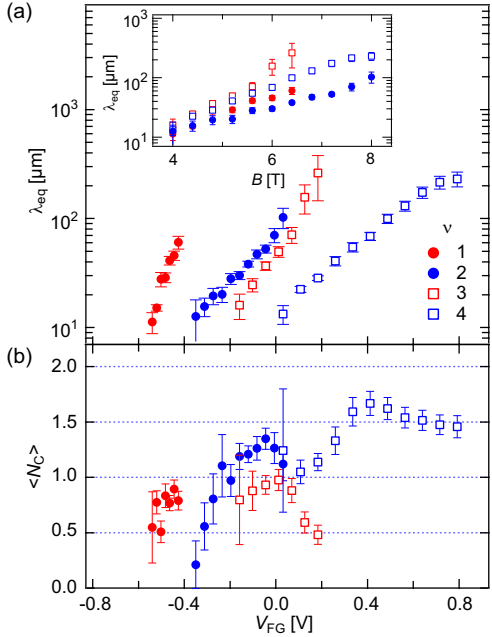}\caption{
(a) Equilibration length $\lambda_\text{eq}$ and (b) effective number of counterflowing modes $\langle N_\text{C}\rangle$ for $\nu = 1$--$4$ obtained by fitting the $g_\text{C}$ vs $L_\text{edge}$ data, plotted as a function of $V_\text{FG}$.
The data in (a) are replotted vs $B$ in the inset.
}
\label{Fig3}
\end{figure}

As Fig.~\ref{Fig2}(c) shows, $g_\text{C}$ decreases with increasing $L_\text{edge}$ for all $V_\text{FG}$.
In the following we restrict our analysis to the $g_\text{C}$ values at integer bulk filling [shown by symbols in Fig.~\ref{Fig2}(c)], where we confirmed the absence of bulk conduction.
In Fig.~\ref{Fig2}(d), we plot $N_\text{C}T_\text{C}$ ($= g_\text{C}$) at $6$~T as a function of $L_\text{edge}$ for $\nu = 1$--4.
The data were then fitted with a single exponential function $N_\text{C} T_\text{C} = A\exp (-L_\text{edge}/\lambda_\text{eq} )$ using $A$ and $\lambda_\text{eq}$ as fitting parameters.
As $T_\text{C} \rightarrow 1$ is expected for $L_\text{edge} \rightarrow 0$, we see that $A = N_\text{C}$.
We therefore use $\langle N_\text{C} \rangle$ instead of $A$ to represent the effective number of counterflowing modes deduced from the fitting.
For $\nu = 4$, we obtain $\langle N_\text{C} \rangle = 1.67$ and $\lambda_\text{eq} = 70~\mu$m at $B = 6$~T.

We performed similar measurements and analysis for a range of magnetic fields ($B = 4$--8 T).
The results are summarized in Fig.~\ref{Fig3}, where $\lambda_\text{eq}$ and $\langle N_\text{C} \rangle$ obtained for $\nu = 1$--$4$ are plotted as a function of $V_\text{FG}$.
For all $\nu$, $\lambda_\text{eq}$ monotonically increases with increasing $V_\text{FG}$ [Fig.~\ref{Fig3}(a)] and hence  $B$ (inset)~\footnote{
For $\nu = 1$ and $3$, only data for $B \leq 6.4$~T, for which good fitting with the exponential function was obtained, are included.
}.
This suggests that the distance between the forward and counterflowing edge channels increases with $B$, which reduces the scattering between them.
At high fields, $\lambda_\text{eq}$ for $\nu = 3$ and $4$ reaches $\sim 200~\mu$m, the value reported in Ref.~\onlinecite{VanWees1995}.
Interestingly, $\langle N_\text{C} \rangle$ increases with $V_\text{FG}$ and peaks out below $1$ for $\nu = 1$ and $3$, whereas it exceeds $1$ and then levels off below $2$ for $\nu = 2$ and $4$ [Fig.~\ref{Fig3}(b)]~\footnote{
Note that {$ \langle N_\text{C} \rangle$} is the effective number of counterflowing modes, not the actual number of counterflowing edge channels determined by the density profile and magnetic field.
In addition, disorder may affect the value of {$ \langle N_\text{C} \rangle$}; potential fluctuation near the edge may cause $N_\text{C}$ to vary between $2$ and $1$ (or between $1$ and $0$) along the edge, making {$ \langle N_\text{C} \rangle$} non-integer.
}.

\begin{figure}[ptb]
\includegraphics{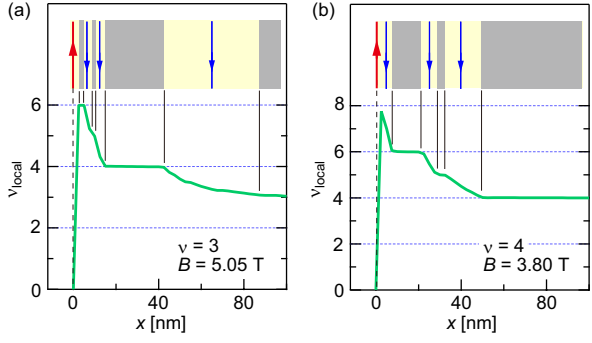}\caption{
Simulated density profile, shown as local filling factor $\nu_\text{local}$, for a bulk electron density of $3.65\times 10^{15}$~m$^{-2}$ at magnetic fields corresponding to bulk filling of (a) $\nu = 3$ and (b) $4$.
The insets are schematics of the top view near the sample edge.
The red (blue) arrows represent forward (counterflowing) edge channels.
Incompressible (compressible) regions are shown in grey (yellow).
}
\label{Fig4}
\end{figure}

To gain insight into the $B$ dependence of $\lambda_\text{eq}$ and the even-odd behavior of $\langle N_\text{C}\rangle$, we simulated the density profile near the mesa edge by solving the Poisson equation self-consistently within the semiclassical approach taking only Landau quantization into account~\footnote{The Poisson equation was solved in the two-dimensional plane perpendicular to the sample edge.
We employed a simplified geometry, with a $400$-nm-wide zero-thickness channel surrounded by $100$-nm-thick insulator with a dielectric constant of $\epsilon_\text{r} = 15$ and a metallic gate in all four directions.
Effective mass of $0.026m_{e}$ and $g$-factor of 10 were used to calculate the energies of the Landau levels, which were then broadened by a Gaussian function with $\sigma = 0.3$~meV.
Finite-temperature effects were not included.
A fixed line charge of $-0.05e$~nm$^{-1}$ along the channel edge was assumed to obtain realistic density profiles.
}.
In Fig.~\ref{Fig4}, we compare the density profiles for (a) $\nu = 3$ and (b) $4$ at the same bulk density of $3.65 \times 10^{15}$~m$^{-2}$.
In both cases, density increases toward the edge, where it drops sharply to zero.
Notably, density varies in a stepwise manner due to the formation of compressible and incompressible strips~\cite{Chklovskii1992}.
As the charge equilibration between adjacent edge channels occurs via scattering across the incompressible strip between them~\cite{Alphenaar1990a, Cui2016}, its width is the important parameter determining the scattering rate.
The width is determined by the density gradient at $B = 0$ and the Landau-level energy separation at the strip~\cite{Chklovskii1992}, the latter being the cyclotron and Zeeman energy for even and odd local filling ($\nu_\text{local}$), respectively.
Our simulations reveal an important role played by the innermost incompressible strip with even $\nu_\text{local}$.
For odd bulk filling $\nu = 3$, the one with $\nu_\text{local} = 4$ is the widest [Fig.~\ref{Fig4}(a)], reflecting the small density gradient (at $B = 0$) and the large cyclotron gap, which then isolates one inner counterflowing channel from all other channels.
The outer counterflowing channels are very close to the forward channels and easily equilibrated with them.
This explains why only one counterflowing mode can transmit for odd $\nu$.
In contrast, for $\nu = 4$, the widest incompressible strip develops at $\nu_\text{local} = 6$ [Fig.~\ref{Fig4}(b)], which isolates two inner counterflowing channels, allowing more than one counterflowing modes to transmit.
The $B$ dependence of $\lambda_\text{eq}$ can be understood in terms of the incompressible-strip width, which we discuss later in detail.

\begin{figure}[ptb]
\includegraphics{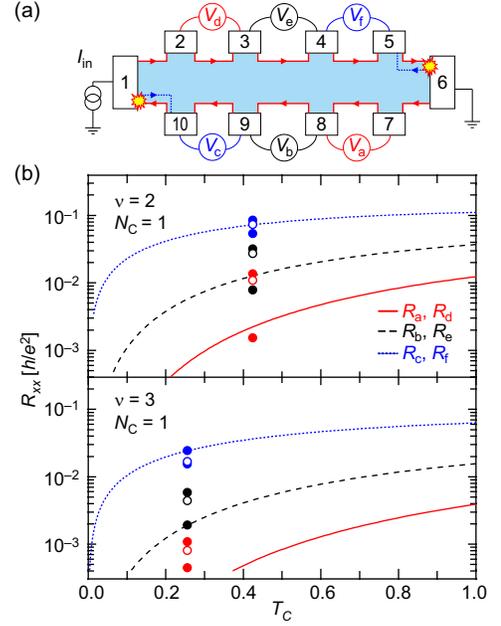}\caption{
(a) Configurations used for the calculation of $R_{xx}$. Yellow markers at the upper-right and lower-left corners represent hot spots.
(b) $R_{xx}$ calculated as a function of $T_\text{C}$ for $\nu = 2$ (upper panel) and $3$ (lower panel).
$R_{xx}$ values for different probes, $R_\alpha = V_\alpha/I_\text{in}$ ($\alpha = \text{a},\ldots,\text{f}$), are shown.
Circles are experimental data in Fig.~\ref{Fig1}, plotted vs $T_\text{C}$ calculated using the $\lambda_\text{eq}$ values in Fig.~\ref{Fig3}(a).
Open circles are data for $B<0$, which are included by using the relation $R_\text{a}(-B) = R_\text{c}(B)$.}
\label{Fig5}
\end{figure}

Now we discuss the probe-position and field-direction dependence of the QH effect breakdown presented in Fig.~\ref{Fig1}.
Using the Landauer-B\"{u}ttiker model, we calculate $R_{xx}$ as a function of $T_\text{C}$ for the configuration shown in Fig.~\ref{Fig5}(a).
The current-voltage relation can be expressed as $\vec{I} = G_{0} \mathbf{M} \vec{V}$, where $\vec{I} = (\cdots, I_i, \cdots)^\mathrm{T}$ and $\vec{V} = (\cdots, V_i, \cdots)^\mathrm{T}$ with $I_i$ ($V_i$) the current (voltage) of the $i$th contact ($i = 1$--$10$)~\cite{Cui2016}.
$\mathbf{M}$ is a matrix with non-zero elements given by
\begin{align*}
& M_{i, i} = (\nu + N_\text{C})T_\text{F}^{(i)} + N_\text{C}T_\text{C}^{(i-1)}\\
& M_{i, i+1} = -N_\text{C} T_\text{C}^{(i)}\\
& M_{i, i-1} =-(\nu+N_\text{C}) T_\text{F}^{(i-1)}
\end{align*}
($i \bmod 10$), where $T_\text{F}^{(i)}$ is the transmission probability of the forward mode on the $i$th edge.
Scattering between forward and counterflowing modes is described by the detailed balance as $(\nu + N_\text{C})(1 - T_\text{F}^{(i)}) = N_\text{C}(1 - T_\text{C}^{(i)})$.
Since all the edges have the same length in the present case, we assume that they share the same $T_\text{C}$ and $T_\text{F}$ values.
We then solved the above equations with $I_{1} = I_\text{in}$, $I_{6} = -I_\text{in}$, and $V_{6} = 0$.

The $T_\text{C}$ dependence of $R_{xx}$ calculated for $\nu = 2$ and $3$ is shown in Fig.~\ref{Fig5}(b).
For these calculations, we took $N_\text{C} = 1$, for comparison with the experiment at $V_\text{FG} = 0$~V [Fig.~\ref{Fig3}(b)].
The experimental data taken from Fig.~\ref{Fig1} are plotted in Fig.~\ref{Fig5}(b) against $T_\text{C}$ [$= \exp(-L_\text{edge}/\lambda_\text{eq})$] calculated using $\lambda_\text{eq}$ for $\nu = 2$ and $3$ at $V_\text{FG} = 0$~V [Fig.~\ref{Fig3}(a)].
Since the model predicts $R_\text{a}(-B) = R_\text{c}(B)$, we included in Fig.~\ref{Fig5}(b) the data for $B < 0$  using this relation.
The calculation reproduces the experimentally observed probe-position dependence, $R_\text{a} < R_\text{b} < R_\text{c}$ for $B > 0$, including the quantitative values.
We note that at $V_\text{FG} = 0$~V the equilibration lengths for $\nu = 2$ and $3$ ($\lambda_\text{eq} \sim 70$ and $50~\mu$m, respectively) are comparable to $L_\text{edge}$ ($= 60~\mu$m).
Hence, the counterflowing mode, being not fully equilibrated with the forward mode, carries charge to the electrode on the upstream and destroys the QH effect.
In contrast, $\lambda_\text{eq} = 13~\mu$m for $\nu = 4$ at $V_\text{FG} = 0$~V is much shorter than $L_\text{edge}$, implying a nearly full equilibration~\footnote{
Indeed, $T_\text{c}=0.01$ for $\nu = 4$ yields $R_{xx} \leq 6 \times 10^{-4}(h/e^{2})$.}.
This explains why the $\nu = 4$ QH effect is well developed at $V_\text{FG} = 0$~V, despite the presence of the counterflowing edge channels.

The probe-position and field-direction dependence can be understood intuitively by considering hot spots~\cite{Klag1991, Komiyama2006}.
In the absence of counterflowing modes, the chemical potential of a forward mode just follows that of the current terminal on its upstream.
Consequently, all the applied bias between the source and drain contacts is concentrated at the two corners where the forward mode meets the source and drain contacts (``hot spots'') [Fig.~\ref{Fig5}(a)].
In contrast, the chemical potential of the counterflowing mode follows primarily that of the electrode on its immediate downstream.
Therefore, the largest chemical potential difference between the forward mode and counterflowing one occurs near the immediate upstream of the hot spots, yielding the chiral QH breakdown behavior.

We now turn to the $B$ dependence of $\lambda_\text{eq}$.
As shown in Fig.~\ref{Fig3}(a), $\lambda_\text{eq}$ increases exponentially with $B$ for both even and odd $\nu$, with nearly the same slope.
The inter-edge-channel scattering rate is governed by the wave function overlap between the states involved, which scales as $\propto (d/\ell_B)^2$, where $d$ is the inter-edge distance and $\ell_B=\sqrt{h/2\pi e B}$ is the magnetic length.
If $d$ is given by the width of the innermost incompressible strip with even $\nu_\text{local}$, it is proportional to the square root of the cyclotron energy~\cite{Chklovskii1992} and hence scales as $\sqrt{B}$.
Since $d/\ell_B \propto B$ in this case, one expects $\lambda_\text{eq}^{-1} \propto \exp [-(B/B_0)^2]$, with $B_0$ a constant~\cite{Martin1990}.
The experimentally observed dependence, $\lambda_\text{eq}^{-1} \propto \exp (-B/B_0)$, is different~\footnote{
Similar exponential $B$ dependence is known for co-propagating edge channels in GaAs, which was observed when $B$ was varied around integer fillings at a fixed density and explained by the increase in the inter-edge distance with decreasing $\nu$~\cite{Chklovskii1992}.
In the present case, $g_\text{C}$ drops more quickly with $L_\text{edge}$ as $\nu$ is slightly reduced from integer values [Fig.~\ref{Fig2}(c)], suggesting that the inter-edge distance becomes smaller with decreasing $\nu$
}, suggesting the relevance of multiple scattering with impurities~\cite{Martin1991}.

Several differences between our results and the previous ones reported for InAs~\cite{VanWees1995} and graphene~\cite{Cui2016} are worth noting.
In Ref.~\onlinecite{VanWees1995}, (i) $N_\text{C}$ was non-zero only for $V_\text{FG} \lesssim -0.4$~V, and (ii) $N_\text{C}$ increased linearly up to $6$ with decreasing $V_\text{FG}$.
In our experiment, $\langle N_\text{C} \rangle$ does not show a monotonic $V_\text{FG}$ dependence, being non-zero for both $V_\text{FG} < 0$ and $V_\text{FG} > 0$, with the maximum value peaked out below $2$.
The chemical properties of the edge~\cite{Mittag2017} and the relative distances of the bulk and edge to the gate~\footnote{
Although not shown here, our experiments suggest that the edge potential depends also on other factors such as the quantum well thickness, distance from the surface, and the history of gate sweep, which will be reported separately.} may partly account for these differences.
However, as our simulations show, the outer counterflowing channels are spatially very close to the forward channels, making it rather unlikely for many of them to transmit~\footnote{
It is not clear whether the large $N_\text{C}$ values found in Ref.~\onlinecite{VanWees1995} originate from the various assumptions made in the analysis.
Our approach, in which one directly measures the counterflowing charge current for each individual edge, provides reliable values for {$\langle N_\text{C} \rangle$} and $T_\text{C}$.
}.
In Ref.~\onlinecite{Cui2016}, despite significant charge accumulation at the edges, QH effects were observed, but at gate voltages shifted from the integer bulk filling.
In the edge-state picture, the transport quantization was explained as resulting from strong scattering between forward and counterflowing channels (i.e., short $\lambda_\text{eq}$) and their isolation from the conductive bulk by the incompressible strip~\footnote{
This happens only when the incompressible strip isolates the bulk from Ohmic contacts.
Whether this happens or not depends on the density profile near Ohmic contacts~\cite{Dahlem2010}.
}.
The microscopic structure of the edge states is non-trivial also in this case, which must be taken into account when making a superconducting junction~\cite{Amet2016,Lee2017}.

In summary, we investigated counterflow edge transport in InAs quantum wells in the QH regime and clarified how it equibrates or manifests as transport anomaly depending on the magnetic field, filling factor, and contact configuration.
Our results suggest that counterflowing edge channels can exist in various systems with sharp edge potential.
Thus, even in the integer QH regime, the microscopic structure of edge states and hence the transport phenomena therein can be more complex than naively expected from the bulk-edge correspondence and should be carefully studied.

The authors thank Yasuhiro Tokura and Masayuki Hashisaka for fruitful discussions and Hiroaki Murofushi for processing the devices. This work was supported by JSPS KAKENHI Grant No. JP15H05854.

\end{document}